\documentclass[12pt]{article}
\usepackage{graphicx}
             
\title{Homogeneous Spaces of the Lorentz Group.}
\author{M. Toller \thanks{e-mail: toller@iol.it}\\ 
via Malfatti n. 8  \\
I-38100 Trento, Italy}

\begin{document} 
\maketitle
                 
\begin{abstract}
We present a classification, up to isomorphisms, of all the homogeneous spaces of the Lorentz group with dimension lower than six. At the same time, we classify, up to conjugation, all the non-discrete closed subgroup of the Lorentz group and all the subalgebras of the Lorentz Lie algebra. We also study the covariant mappings between some pairs of homogeneous spaces. This exercise is done without any claim of originality, but with the hope of providing a useful instrument for the investigation of the Lorentz transformation properties of some physical quantities.
              
\end{abstract}

\renewcommand{\labelenumii}{\Alph{enumii}.}

\section{Introduction.}

In the last few years there has been a considerable interest in a possible failure of the Lorentz symmetry, suggested both by theoretical speculations and by experimental observations.  It has also been proposed \cite{AC,MS,KG} that the relativity principle and the Lorentz symmetry are valid, but the action of the Lorentz group on the components of the energy-momentum is ``deformed'' and is not linear any more. More references and a more detailed discussion are given in ref.\ \cite{Toller}.

In the present paper we present some mathematical tools useful for the discussion of the above mentioned problems. They are simple mathematical exercises, but it is not easy to find an explicit and compact treatment of them in the literature. The physical relevance of some results is discussed in a shorter forthcoming paper \cite{Toller2}.

It is convenient to consider a set of classical physical observables as continuous coordinates in an abstract topological space $\Pi$. This is also possible in a quantum treatment if the observables are compatible. They are described by commuting self-adjoint operators which generate an abelian $C^*$ algebra, which, according to a theorem by Gelfand \cite{Naimark}, is isomorphic to an algebra of continuous functions defined on a locally compact space $\Pi$. Also in the classical case the observables generate an abelian $C^*$ algebra, and we assume in general that $\Pi$ is locally compact.

The symmetry group $\mathcal{G}$, which is a Lie group \cite{Chevalley} or, at least, a topological group \cite{Pontryagin}, acts continuously on $\Pi$ and we indicate this action by the notation $(g, \pi) \to g\pi$, where  $g \in \mathcal{G}$ and  $\pi \in \Pi$. Of course, we have $g(g'\pi) = (gg')\pi$. We are not assuming that $\Pi$ is a vector space and it is not even meaningful to say that the action of  $\mathcal{G}$ is or is not linear.

We begin by recalling some important definitions and results. If there is another space  $\Pi'$ on which $\mathcal{G}$ acts continuously, a continuous mapping $\rho: \Pi \to \Pi'$ is called {\it covariant} (or equivariant) if it commutes with the action of the group, namely if $\rho(g \pi) =  g \rho(\pi)$.  The action of $\mathcal{G}$ defines on $\Pi$ an equivalence relation and the equivalence classes are the {\it orbits}. Two points $\pi, \pi'$ belong to the same orbit if $\pi' = g \pi$ with $g \in \mathcal{G}$. 

If $\Pi$ is composed of a single orbit, we say that the action of $\mathcal{G}$ is {\it transitive}. This happens in many physical problems, in particular the problem which has motivated this research. In the other cases it is always possible to decompose $\Pi$ into orbits on which $\mathcal{G}$ acts transitively. In the following we assume that the action of $\mathcal{G}$ on $\Pi$ is transitive. 

We choose an element $\pi \in \Pi$ and we indicate by  $\mathcal{H}$ the corresponding {\it stabilizer subgroup} defined by the condition $g\pi = \pi$. It is a closed subgroup of $\mathcal{G}$. We see immediately that there is a one-to-one correspondence between the points of $\Pi$ and the cosets which form the quotient space $\mathcal{G}/\mathcal{H}$. One defines in a natural way a topology and a continuous action of $\mathcal{G}$ on this space which it is called a (topological) {\it homogeneous space}. 

The bijective mapping $\mathcal{G}/\mathcal{H} \to \Pi$ is continuous and covariant, but the inverse mapping is not necessarily continuous. In order to simplify the discussion, we replace the topology of $\Pi$ by a finer topology in such a way that $\mathcal{G}/\mathcal{H}$ and $\Pi$ are omeomorphic. Physically, this means to admit a larger class of continuous observables. With this topology, $\Pi$ too is called a topological homogeneous space. Note that if $\mathcal{G}$ is locally compact, $\Pi$ is locally compact too, in agreement with the assumption justified above. 

We say that two homogeneous spaces (with respect to the same group) are {\it isomorphic} if they are connected by an covariant continuous mapping with continuous inverse. In particular, $\Pi$ is isomorphic to $\mathcal{G}/\mathcal{H}$. If $\pi' = g \pi$, its stabilizer is  $\mathcal{H}' = g \mathcal{H} g^{-1}$, namely the two stabilizers are {\it conjugated}. The homogeneous spaces $\mathcal{G}/\mathcal{H}$ and $\mathcal{G}/\mathcal{H}'$ are isomorphic and it follows that, in order to classify all the homogeneous spaces of $\mathcal{G}$ up to isomorphisms, it is sufficient to classify all its closed subgroups up to conjugation. 

In our case, $\mathcal{G}$  is a Lie group and its closed subgroups, in particular the stabilizer $\mathcal{H}$, are Lie groups too \cite{Bourbaki}. The homogeneous space $\mathcal{G}/\mathcal{H}$ has a structure of analytic manifold, which can be transmitted to $\Pi$. Note that if the subgroups $\mathcal{H}'$ and $\mathcal{H}$ are conjugate, namely $\mathcal{H}' = g \mathcal{H} g^{-1}$, the corresponding Lie subalgebras $\mathbf{h}'$ and $\mathbf{h}$ are conjugated too, namely we have $\mathbf{h}' = ad_g \mathbf{h}$, where $g \to ad_g$ is the adjoint representation of $\mathcal{G}$.

Then, the first step is to classify all the subalgebras of the Lie algebra $\mathbf{g}$ of $\mathcal{G}$ up to conjugation. Each Lie subalgebra $\mathbf{h}$ determines uniquely a connected Lie subgroup $\mathcal{H}$ and the most general Lie subgroup is obtained by adding to this connected subgroup other connected components.

In the following sections we carry out this program for the group $\mathcal{G} = SL(2, \mathbf{C})$ (considered as a real Lie group), locally isomorphic to the proper orthochronous Lorentz group $SO^{\uparrow}(1, 3)$.  We note that an homogeneous space of $SL(2, \mathbf{C})$ is also an homogeneous space of $SO^{\uparrow}(1, 3)$ if the stabiblizer $\mathcal{H}$ contains the central element
\begin{equation}
-e = \left( \begin{array}{cc}
-1 & 0 \\ 0 & -1
\end{array} \right).
\end{equation}
  
\section{Lie subalgebras.}

Now we have to classify, up to conjugation, all the Lie subalgebras of  $\mathbf{g} = sl(2, \mathbf{C}) = o(1, 3)$. We introduce a basis composed of the generators of the rotations $M_r$ and the generators of the Lorentz boosts $L_r$ ($r = 1, 2, 3)$, which satisfy the familiar commutation relations
\begin{equation}
[M_r, M_s] = \epsilon_{rst} M_t, \qquad
[M_r, L_s] = \epsilon_{rst} L_t, \qquad
[L_r, L_s] = - \epsilon_{rst} M_t.
\end{equation}
A generic element of $\mathbf{g}$ can be written in the form
\begin{equation}
A = \alpha_r  M_r + \beta_r  L_r
 = \vec \alpha  \cdot \vec M + \vec \beta  \cdot \vec L. 
\end{equation}

The adjoint representation of $\mathcal{G} = SL(2, \mathbf{C})$ is equivalent to the representation which acts on the antisymmetric second order tensors. It acts on the Lie algebra $\mathbf{g}$, which can be decomposed into orbits of conjugate elements. The following two quantities are Lorentz-invariant and are therefore constant on the orbits
\begin{equation} 
c_1 = |\vec\alpha|^2 - |\vec\beta|^2, \qquad c_2 = \vec\alpha \cdot \vec\beta.
\end{equation}

It is easy to show that every non-vanishing element of $\mathbf{g}$ can be Lorentz transformed into one of the following forms, which provide representative elements of the orbits:
\begin{equation}
A = \mu M_3 + \nu L_3, \qquad \mu > 0, \quad \nu \neq 0, \qquad
c_1 = \mu^2 - \nu^2, \quad c_2 = \mu\nu, 
\end{equation}
\begin{equation}
A = \mu M_3 \qquad \mu > 0, \qquad
c_1 =  \mu^2, \quad c_2 = 0,
\end{equation}
\begin{equation}
A = \nu L_3 \qquad \nu > 0, \qquad
c_1 = - \nu^2, \quad c_2 = 0,
\end{equation}
\begin{equation} \label{nilpotent}
A = M_1 + L_2, \qquad c_1 = c_2 = 0.
\end{equation}
The elements belonging to the last orbit, defined by $c_1 = c_2 = 0$, are called {\it nilpotent}, because some power of the corresponding operator of the adjoint representation vanishes. The other elements are called {\it semisimple}, because, as we shall see, the corresponding operators of the adjoint representation are semisimple. All these orbits are four-dimensional and we shall describe them with more detail in section 4.

We also obtain the following classification, up to conjugation, of the zero and one-dimensional subalgebras:
\begin{enumerate}
\item $\mathbf{h}_6$, containing only the zero element;
\item $\mathbf{h}_5^{\lambda}$, ($\lambda \neq 0$), generated by the element $M_3 + \lambda L_3$;
\item $\mathbf{h}_5^0$, generated by the element $M_3$;
\item $\mathbf{h}_5^{\infty}$, generated by the element $L_3$;
\item $\mathbf{h}_5^N$, generated by the nilpotent element $M_1 + L_2$.
\end{enumerate}
The subscript is the codimension of the algebra, equal to the dimension of the corresponding homogeneous space.

The next step is to find (up to conjugation) all the subalgebras composed exclusively of nilpotent elements, besides $\mathbf{h}_5^N$. We may assume that one element is given by $M_1 + L_2$ and we write the generic element in the form
\begin{equation} 
A = \mu(M_1 + L_2)  + \vec \alpha  \cdot \vec M + \vec \beta  \cdot \vec L. 
\end{equation}
This element must be nilpotent for any value of $\mu$ and it follows that
\begin{equation} 
|\vec\alpha|^2 - |\vec\beta|^2 = \vec\alpha \cdot \vec\beta = 0,
\end{equation}
\begin{equation} 
\alpha_1 - \beta_2 = 0, \qquad \alpha_2 + \beta_1 = 0. 
\end{equation}
With some calculations, we obtain
\begin{equation} 
\alpha_3 = \beta_3 = 0 
\end{equation}
and we have
\begin{equation} 
A = (\mu + \alpha_1)(M_1 + L_2) + \alpha_2(M_2 - L_1).
\end{equation}
These elements form a commutative two-dimensional Lie subalgebra, which we add to our list:
\begin{enumerate} \setcounter{enumi}{5}
\item $\mathbf{h}_4^N$, generated by the elements $M_1 + L_2$ and $M_2 - L_1$.
\end{enumerate}

All the other subalgebras contain at least a semisimple element and we may assume that it has the form $\mu M_3 + \nu L_3$. We consider the corresponding operator $\rho$ of the adjoint representation defined by
\begin{equation} 
\rho A = [\mu M_3 + \nu L_3, A], \qquad A \in \mathbf{g}. 
\end{equation}
As we have anticipated above, this operator is semisimple, namely we can introduce a basis in the complexification $\mathbf{g}_C$ of $\mathbf{g}$ in which it is diagonal. In fact, we have
\begin{displaymath} 
\rho M_3 = 0,
\end{displaymath}
\begin{displaymath} 
\rho L_3 = 0,
\end{displaymath}
\begin{displaymath} 
(\rho - i \mu - \nu)(M_1 + L_2 - i M_2 + i L_1) = 0,
\end{displaymath}
\begin{displaymath} 
(\rho + i \mu - \nu)(M_1 + L_2 + i M_2 - i L_1) = 0,
\end{displaymath}
\begin{displaymath} 
(\rho - i \mu + \nu)(M_1 - L_2 - i M_2 - i L_1) = 0,
\end{displaymath}
\begin{equation} \label{Eigen}
(\rho + i \mu + \nu)(M_1 - L_2 + i M_2 + i L_1) = 0.
\end{equation}

The subspace $\mathbf{h}$ and its complexification $\mathbf{h}_C$ must be invariant under the operator $\rho$ and in $\mathbf{h}_C$ there is a basis composed of eigenvectors of $\rho$, chosen among the eigenvectors listed above. Since we want to build a real basis, every complex eigenvector must be chosen together with its complex conjugate. If there are degenerate eigenvalues, a linear superposition of the corresponding eigenvectors can be chosen. In this way we easily classify all the invariant subspaces. Then we have to check if they are subalgebras.

We consider first the case in which $\mathbf{h}$ contains both the elements $M_3$ and $L_3$ and therefore also $\mu M_3 + \nu L_3$ with an arbitrary choice of the coefficients. The complex eigenvalues are not degenerate and we may choose zero, one or two pairs of complex eigenvectors, obtaining the subalgebras
\begin{enumerate} \setcounter{enumi}{6}
\item $\mathbf{h}_4$, generated by the elements $M_3$ and $L_3$;
\item $\mathbf{h}_2$, generated by the elements $M_3$, $L_3$, $M_1 + L_2$ and $M_2 - L_1$;
\item $\mathbf{h}_0 = sl(2, \mathbf{C})$.
\end{enumerate}
By choosing the second pair of complex eigenvectors instead of the first, we obtain a conjugate subalgebra.

Then we assume that only the linear superposition $\mu M_3 + \nu L_3$ belongs to $\mathbf{h}$ and that the coefficients do not vanish. We may assume $\mu = 1$ and $\nu = \lambda \neq 0$. In this case too, the complex eigenvalues are not degenerate and we may choose one or two pairs of complex eigenvectors. In the second case, the linear subspace is not a subalgebra and in the first case we obtain
\begin{enumerate} \setcounter{enumi}{9}
\item $\mathbf{h}_3^{\lambda}$, ($\lambda \neq 0$), generated by the elements $M_3 + \lambda L_3$, $M_1 + L_2$ and $M_2 - L_1$.
\end{enumerate}
In this case too, choosing the second pair of complex eigenvectors instead of the first, we obtain a conjugate subalgebra.

Another possibility is that $\mu = 1$, $\nu = 0$ and $L_3 \notin \mathbf{h}$. Then the pairs of complex eigenvalues are degenerate and one can choose a pair of eigenvectors belongoing to two-dimensional vector spaces, namely
\begin{equation} 
\alpha (M_1 - i M_2) + \beta (L_2 + i L_1)
\end{equation}
and its complex conjugate. These vector spaces cannot be entirely contained in $\mathbf{h}$, otherwise we obtain  $L_3 \in \mathbf{h}$. For the same reason, we must have $\alpha \overline\beta = \overline\alpha \beta$ and we may assume that $\alpha$ and $\beta$ are real.

This expression can be simplified by means of a boost generated by $L_3$, which acts in the following way:
\begin{displaymath} 
M_3 \to M_3, \qquad L_3 \to L_3, 
\end{displaymath}
\begin{displaymath} 
M_1 - i M_2 \to \cosh\zeta \, (M_1 - i M_2) + \sinh\zeta \, (L_2 + iL_1), 
\end{displaymath}
\begin{equation} 
L_2 + iL_1\to \cosh\zeta \, (L_2 + iL_1) + \sinh\zeta \, (M_1 - iM_2).
\end{equation}
This transformation permits us to consider only three special cases, namely $\beta = 0$, $\alpha = 0$ or $\alpha = \beta$.

We obtain in this way the following subalgebras, which correspond to the {\it little groups} introduced by Wigner \cite{Wigner}. 
\begin{enumerate} \setcounter{enumi}{10}
\item $\mathbf{h}_3^+ = su(2) = o(3)$, generated by the elements $M_3$,  $M_1$ and $M_2$;
\item $\mathbf{h}_3^- = su(1, 1) = o(1, 2)$, generated by the elements $M_3$, $L_1$, and $L_2$;
\item $\mathbf{h}_3^0 = e(2)$, generated by the elements $M_3$, $M_1 + L_2$, and $M_2 - L_1$.
\end{enumerate}

The last possibility is that $\mu = 0$, $\nu = 1$ and $M_3 \notin \mathbf{h}$. Since the eigenvalues of $\rho$ are real, it is not necessary to introduce the complexification of $\mathbf{g}$ and to choose pairs of conjugate eigenvectors. The eigenvalues are degenerate and we have to consider the following two-dimensional spaces of eigenvectors 
\begin{displaymath} 
\alpha (M_1 + L_2) +\beta (M_2 - L_1),
\end{displaymath}
\begin{equation} 
\gamma (M_1 - L_2) + \delta (M_2 + L_1).
\end{equation}
In order to avoid the appearance of $M_3$ in $\mathbf{h}$, we must have $\alpha \delta = \beta \gamma$. This equality implies that one cannot choose more that two independent eigenvectors. We may choose two independent eigenvectors in the first vector space or only one of them, which, by means of a rotation generated by $M_3$, can be transformed into an element proportial to $M_1 + L_2$. We obtain in this way the following subalgebras
\begin{enumerate} \setcounter{enumi}{13}
\item $\mathbf{h}_4^{\infty}$, generated by the elements $L_3$ and $M_1 + L_2$;
\item $\mathbf{h}_3^{\infty}$, generated by the elements $L_3$, $M_1 + L_2$ and $M_2 - L_1$.
\end{enumerate}
By choosing one or two elements in the second vector space, we obtain conjugate subalgebras. We may also choose  an element in the first and another element in the second vector space and by means of a rotation we can set $\beta = 0$. The constraint given above implies $\delta = 0$ and we obtain a subalgebra generated by $L_3$, $M_1$ and $L_2$, conjugate to the subalgebra $\mathbf{h}_3^-$, which has already been considered. 

\begin{figure}[t]
\includegraphics[width=13.5cm]{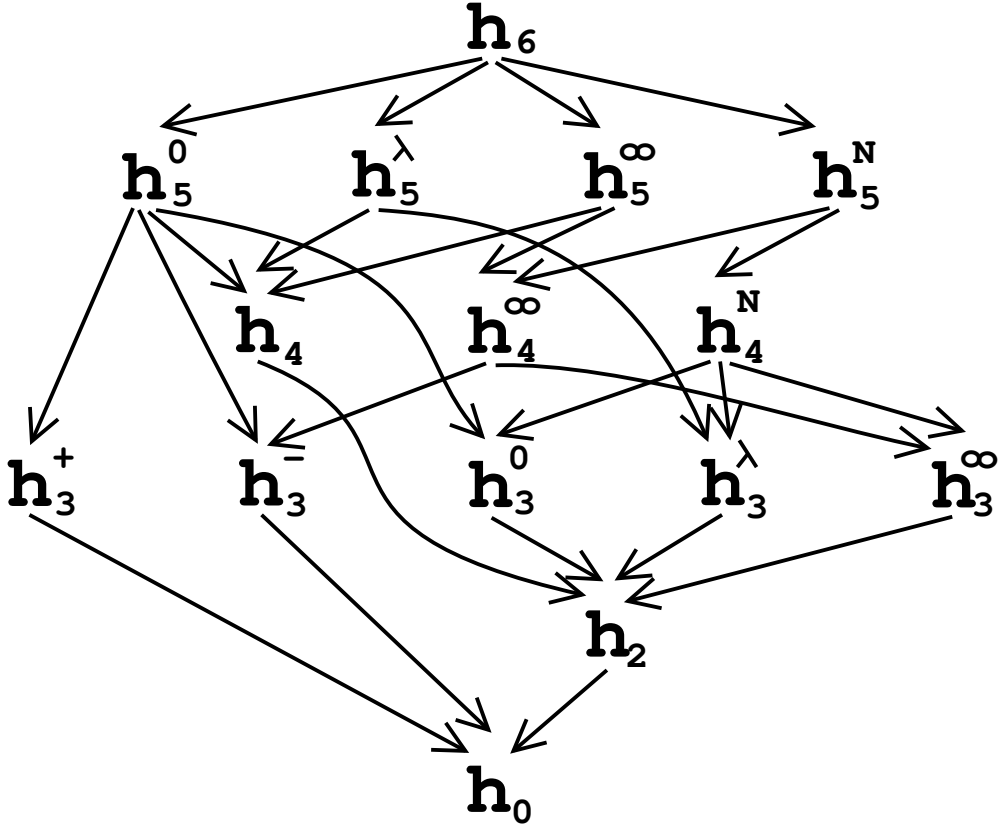}
\caption{Subalgebras of $sl(2, \mathbf{C})$.}
\end{figure}

In conclusion, all the subalgebras are conjugate to one of the subalgebras described above. They are listed in a more natural order in Figure 1.  The arrow $\mathbf{h} \to \mathbf{h}'$ means that $\mathbf{h}'$ contains a subalgebra conjugate to $\mathbf{h}$. 

Note that $\mathbf{h}_5^0$ and $\mathbf{h}_5^{\infty}$ can naturally be considered as limits of the family  $\mathbf{h}_5^{\lambda}$ and, in a similar way,  $\mathbf{h}_3^0$ and $\mathbf{h}_3^{\infty}$ are limits of the family $\mathbf{h}_3^{\lambda}$. We have listed them separately because they have some different features. There is no subalgebra of dimension five, and therefore there is no homogeneous space of dimension one. Only the subalgebras $\mathbf{h}_0$, $\mathbf{h}_3^+$ and $\mathbf{h}_3^-$ are semisimple. The subalgebras $\mathbf{h}_0$, $\mathbf{h}_2$, $\mathbf{h}_4$, $\mathbf{h}_4^N$ and $\mathbf{h}_6$ can be considered as complex subalgebras of $sl(2, \mathbf{C})$. 

\section{Closed subgroups.}

We describe the closed subgroups of $SL(2, \mathbf{C})$ by giving explicitly the corresponding $2 \times 2$ complex matrices. We indicate by $\mathbf{Z}$, $\mathbf{R}$ and $\mathbf{C}$ the additive groups of the integers, the real and the complex numbers and by $\mathbf{R}^*$ and $\mathbf{C}^*$ the multiplicative groups of the non-vanishing real and complex numbers. In terms of matrices, the basis elements of the Lie algebra $sl(2, \mathbf{C})$ are given by
\begin{displaymath}
L_1 = \sigma_1 = \left( \begin{array}{cc}
0 & 1 \\ 1 & 0
\end{array} \right),
\end{displaymath}
\begin{displaymath} 
L_2 = \sigma_2 = \left( \begin{array}{cc}
0 & -i \\ i & 0
\end{array} \right),
\end{displaymath}
\begin{displaymath} 
L_3 = \sigma_3 = \left( \begin{array}{cc}
1 & 0 \\ 0 & -1
\end{array} \right), 
\end{displaymath}
\begin{equation}
M_1 = -i \sigma_1, \qquad M_2 = -i \sigma_2, \qquad M_3 = -i \sigma_3.
\end{equation}
The various connected Lie subgroups are generated by the exponentials of the elements of the corresponding Lie subalgebras and are labelled in a similar way. In all the cases we are considering, the subgroups obtained in this way are closed.

All the other closed subgroups are obtained by adding new connected components. If $\mathcal{H}$ is the connected component of the identity of the subgroup $\mathcal{H}'$, an element $g \in \mathcal{H}'$ has the property $g \mathcal{H} g^{-1} = \mathcal{H}$, namely it belongs to the {\it normalizer} $\mathcal{N}$ of $\mathcal{H}$. $\mathcal{H}$ is a normal subgroup of $\mathcal{N}$  and the connected components of $\mathcal{H}'$ are cosets of $\mathcal{H}$ belonging to a discrete subgroup of the quotient group $\mathcal{N}/\mathcal{H}$. We describe the elements of $\mathcal{N}/\mathcal{H}$ by means of a set $\mathcal{Q}$ of representative elements chosen in the cosets. Sometimes, but not always, it is possible to choose these representative elements in such a way that $\mathcal{Q}$ is a subgroup. 

From the list of Lie subalgebras given in the preceding section, we obtain the following list of all the connected closed subgroups, labelled by the same indices. In each case we describe the normalizer $\mathcal{N}$, the quotient group $\mathcal{N}/\mathcal{H}$  and we give the list of all its discrete subgroups, describing them by means of representative elements of the cosets, belonging to a discrete subset of $\mathcal{Q}$. If $g$ is a representative element, the corresponding connected component is $g\mathcal{H} = \mathcal{H} g$. We say that a non-connected subgroup is defined by the corresponding subset of $\mathcal{Q}$. 

In this way we find the list of all the non-connected closed subgroups, which are denoted by the symbols representing their connected components of the identity with some additional subscripts. Details of the calculation of some normalizer subgroups ond of some discrete subgroups are given in the appendices.

\begin{enumerate}

\item $\mathcal{H}_6$, containing only the unit element. The additional connected components are isolated points, which form an arbitrary discrete subgroup. As it is stated in ref.\ \cite{Winkelmann}, the class of the discrete subgroups  of $SL(2, \mathbf{C})$ is (at least to present time) too large for a fine classification.

\item $\mathcal{H}_5^{\lambda}$ ($\lambda \neq 0$), composed of the matrices of the kind
\begin{equation} 
\left( \begin{array}{cc}
\exp(\lambda t - it) & 0 \\ 0 & \exp(-\lambda t + it)) 
\end{array} \right), 
\qquad  t \in \mathbf{R}. 
\end{equation}
The normalizer is given by eq.\ (\ref{Normal}) of the appendix A.2 and the elements of the subgroup $\mathcal{Q}$ have one of the forms
\begin{displaymath} 
\left( \begin{array}{cc}
\exp(i\phi) & 0 \\ 0 & \exp(-i\phi)
\end{array} \right), \qquad
\left( \begin{array}{cc}
0 & -\exp(-i\phi) \\ \exp(i\phi) & 0
\end{array} \right), 
\end{displaymath}
\begin{equation} \label{Q1}
0 \leq \phi < 2\pi.
\end{equation}
The subgroups obtained by adding other connected components are
\begin{enumerate}
\item $\mathcal{H}_{5,n}^{\lambda}$ ($n = 2, 3, 4,\ldots$), defined by the following elements of $\mathcal{Q}$ 
\begin{equation}  \label{Q2}
\left( \begin{array}{cc}
\exp(2i\pi \frac \nu n ) & 0 \\ 0 & \exp(-2i\pi \frac \nu n ) 
\end{array} \right),
\qquad \nu = 0, 1,\ldots, n-1.
\end{equation}
\item $\mathcal{H}_{5,n,\eta}^{\lambda}$ ($n = 2, 4, 6,\cdots, \quad 0 \leq \eta < 1$), defined by the elements (\ref{Q2}) and
\begin{displaymath} 
\left( \begin{array}{cc}
0 & -\exp(- 2i\pi\frac {\nu + \eta} n ) \\ \exp(2i\pi\frac {\nu + \eta} n) & 0 
\end{array} \right),
\end{displaymath} 
\begin{equation} \label{Q3}
\qquad \nu = 0, 1,\ldots, n-1.
\end{equation}
\end{enumerate}

\item $\mathcal{H}_5^0$,  composed of the matrices of the kind
\begin{equation} \label{Q8}
\left( \begin{array}{cc}
\exp(-i\phi) & 0 \\ 0 & \exp(i\phi) 
\end{array} \right), 
\qquad  0 \leq \phi < 2\pi. 
\end{equation}
The normalizer is given by eq.\ (\ref{Normal}) of the appendix A.2 and the elements of $\mathcal{Q}$, which is not a group, have one of the forms
\begin{equation}
\left( \begin{array}{cc}
s & 0 \\ 0 & s^{-1}
\end{array} \right), \qquad
\left( \begin{array}{cc}
0 & -s^{-1} \\ s & 0
\end{array} \right), 
\qquad s > 0.
\end{equation}
The subgroups obtained by adding other connected components are
\begin{enumerate}
\item $\mathcal{H}_{5,k}^0$ ($k > 1$), defined by the elements
\begin{equation} \label{Q4}
\left( \begin{array}{cc}
k^m & 0 \\ 0 & k^{-m}
\end{array} \right),
\qquad m \in \mathbf{Z}.
\end{equation}
\item $\mathcal{H}_{5,k,h}^0$ ($k > h \geq 1$), defined by the elements (\ref{Q4}) and
\begin{equation} 
\left( \begin{array}{cc}
0 & - h^{-1} k^{-m}  \\ h k^m  & 0
\end{array} \right),
\qquad  \qquad m \in \mathbf{Z}.
\end{equation}
\item $\mathcal{H}_{5,1,h}^0$ ($h > 0$), defined by the additional element
\begin{equation} 
\left( \begin{array}{cc}
0 & - h^{-1} \\ h  & 0
\end{array} \right).
\end{equation}
\end{enumerate}

\item $\mathcal{H}_5^{\infty}$,  composed of the matrices of the kind
\begin{equation} \label{Q6}
\left( \begin{array}{cc}
\exp(t) & 0 \\ 0 & \exp(-t) 
\end{array} \right),
\qquad  t \in \mathbf{R}. 
\end{equation}
The normalizer is given by eq.\ (\ref{Normal}) of the appendix A.2 and $\mathcal{Q}$ is defined by eq. (\ref{Q1}). The  subgroups obtained by adding other connected components are
\begin{enumerate}
\item $\mathcal{H}_{5,n}^{\infty}$  ($n = 2, 3, 4,\ldots$), defined by the elements (\ref{Q2}). 
\item $\mathcal{H}_{5,n,\eta}^{\infty}$ ($n = 2, 4, 6,\cdots, \quad 0 \leq \eta < 1$), defined by the elements (\ref{Q2}) and (\ref{Q3}). 
\end{enumerate}

\item $\mathcal{H}_5^N$,  composed of the matrices of the kind
\begin{equation} 
\left( \begin{array}{cc}
1 & is \\ 0 & 1 
\end{array} \right),
\qquad  s \in \mathbf{R}. 
\end{equation}
Following the procedure indicated in the appendix A.3, we see that the normalizer is composed of the matrices
\begin{equation} \label{Q7}
\left( \begin{array}{cc}
i^{\nu}a & b \\ 0 & i^{-\nu}a^{-1} 
\end{array} \right),
\qquad \nu = 0, 1, 2, 3, \quad a > 0, \quad b \in \mathbf{C} 
\end{equation}
and the subgroup $\mathcal{Q}$ contains the matrices
\begin{equation} 
\left( \begin{array}{cc}
i^{\nu}a & i^{\nu}b \\ 0 & i^{-\nu}a^{-1} 
\end{array} \right),
\qquad \nu = 0, 1, 2, 3, \quad a > 0, \quad b \in \mathbf{R}. 
\end{equation}
The discrete subgroups of $\mathcal{Q}$ which have some elements with $a \neq 1$, can be treated by means of the results of appendix A.4 and we find the following subgroups obtained by adding other connected components 
\begin{enumerate} 
\item $\mathcal{H}_{5,k,h,\nu}^N$ ($k > 1$, $h \in \mathbf{R}$, $\nu = 0, 1, 2, 3$), defined by the elements
\begin{equation} 
\left( \begin{array}{cc}
(i^{\nu}k)^m & ((i^{\nu}k)^m - (i^{\nu}k)^{-m})h \\ 0 & (i^{\nu}k)^{-m} 
\end{array} \right), \qquad
m \in \mathbf{Z}.
\end{equation}
\item $\mathcal{H}_{5,k,h,\nu +}^N$ ($k > 1$, $h \in \mathbf{R}$, $\nu = 0, 1$), defined by the elements
\begin{equation}
\pm \left( \begin{array}{cc}
(i^{\nu}k)^m & ((i^{\nu}k)^m - (i^{\nu}k)^{-m})h \\ 0 & (i^{\nu}k)^{-m} 
\end{array} \right), \qquad
m \in \mathbf{Z}.
\end{equation}
\item $\mathcal{H}_{5,k,h ++}^N$ ($k > 1$, $h \in \mathbf{R}$), defined by the elements
\begin{equation}
\left( \begin{array}{cc}
i^{\nu} k^m & (i^\nu k^m - i^{-\nu}k^{-m})h \\ 0 & i^{-\nu} k^{-m} 
\end{array} \right), \qquad
m \in \mathbf{Z}, \quad \nu = 0, 1, 2, 3.
\end{equation}
\end{enumerate}
We have also to consider the subgroups of $\mathcal{Q}$ composed of matrices with $a = 1$. The corresponding subgroups obtained by adding other connected components are
\begin{enumerate} \setcounter{enumii}{3}
\item $\mathcal{H}_{5,1,h,0}^N$ ($h > 0$), defined by the elements
\begin{equation} 
\left( \begin{array}{cc}
1 & mh \\ 0 & 1 
\end{array} \right), \qquad
m \in \mathbf{Z}.
\end{equation}
\item $\mathcal{H}_{5,1,h,2}^N$ ($h > 0$), defined by the elements
\begin{equation} 
(-1)^m \left( \begin{array}{cc}
1 & mh \\ 0 & 1 
\end{array} \right), \qquad
m \in \mathbf{Z}.
\end{equation}
\item $\mathcal{H}_{5,1,h,0+}^N$ ($h \geq 0$), defined by the elements
\begin{equation} 
\pm\left( \begin{array}{cc}
1 & mh \\ 0 & 1 
\end{array} \right), \qquad
m \in \mathbf{Z}.
\end{equation}
\item $\mathcal{H}_{5,1,h,1+}^N$ ($h > 0$), defined by the elements
\begin{equation} 
\pm\left( \begin{array}{cc}
i^m & i^m mh \\ 0 & i^{-m} 
\end{array} \right), \qquad
m \in \mathbf{Z}.
\end{equation}
\item $\mathcal{H}_{5,1,h}^N$ ($h \geq 0$), defined by the four elements $\pm e$ and
\begin{equation} 
\pm\left( \begin{array}{cc}
i & i h \\ 0 & -i 
\end{array} \right).
\end{equation}
\item $\mathcal{H}_{5,1,h++}^N$ ($h > 0$), defined by the elements
\begin{equation} 
\left( \begin{array}{cc}
i^{\nu} & i^{\nu}mh \\ 0 & i^{-\nu} 
\end{array} \right), \qquad
m \in \mathbf{Z}, \quad \nu = 0, 1, 2, 3.
\end{equation}
\end{enumerate}

\item $\mathcal{H}_4^N$, composed of the matrices of the kind
\begin{equation} 
\left( \begin{array}{cc}
1 & z \\ 0 & 1 
\end{array} \right),
\qquad  z \in \mathbf{C}. 
\end{equation}
The normalizer is given by eq.\ (\ref{Triangular}) of the appendix A.3 and the subgroup $\mathcal{Q}$ contains the matrices
\begin{equation} 
\left( \begin{array}{cc}
a & 0 \\ 0 & a^{-1} 
\end{array} \right),
\qquad  a \in \mathbf{C}^*. 
\end{equation}
The subgroups obtained by adding other connected components are
\begin{enumerate}
\item $\mathcal{H}_{4,n}^N$ ($n = 2, 3, 4,\ldots$), defined by the elements (\ref{Q2}),
\item $\mathcal{H}_{4,k,n,\eta}^N$ ($k > 1$, $n = 1, 2, 3,\ldots$, $0 \leq \eta < 1$), defined by the elements
\begin{displaymath} 
\left( \begin{array}{cc}
k^m \exp(2\pi i \frac {\nu + m \eta} n) & 0 \\ 0 & k^{-m} \exp(-2\pi i \frac {\nu + m \eta} n)
\end{array} \right), 
\end{displaymath} 
\begin{equation} 
\qquad  m \in \mathbf{Z},\qquad \nu = 0, 1,\ldots, n - 1.
\end{equation}
\end{enumerate}

\item $\mathcal{H}_4$, composed of the matrices of the kind
\begin{equation} 
\left( \begin{array}{cc}
z & 0 \\ 0 & z^{-1} 
\end{array} \right),
\qquad  z \in \mathbf{C}^*.  
\end{equation}
The normalizer is given by eq.\ (\ref{Normal}) of the appendix A.2 and it is the union of $\mathcal{H}_4$ and $i\sigma_2 \mathcal{H}_4$. Then we have only to consider the non-connected subgroup
\begin{enumerate}
\item $\mathcal{H}_{4,+}$, defined by the additional element $i\sigma_2$.
\end{enumerate}

\item $\mathcal{H}_2$, composed of the complex triangular matrices of the kind (\ref{Triangular}) of the appendix A.3. It coincides with its normalizer and it is not possible to add other connected components.

\item $\mathcal{H}_0 = SL(2, \mathbf{C})$.

\item $\mathcal{H}_3^{\lambda}$ ($\lambda \neq 0$), composed of the matrices of the kind
\begin{equation} 
\left( \begin{array}{cc}
\exp(\lambda t - it) & z \\ 0 & \exp(-\lambda t + it) 
\end{array} \right),
\qquad  t \in \mathbf{R}, \quad z \in \mathbf{C}.  
\end{equation}
The normalizer is given by eq.\ (\ref{Triangular}) of the appendix A.3 and the elements of $\mathcal{Q}$ have the form
(\ref{Q8}). The subgroups obtained by adding other connected components are
\begin{enumerate}
\item $\mathcal{H}_{3,n}^{\lambda}$ ($n = 2, 3, 4,\ldots$), defined by the elements (\ref{Q2}).
\end{enumerate}

\item $\mathcal{H}_3^+ = SU(2)$, composed of the matrices of the kind
\begin{equation} 
\left( \begin{array}{cc}
\alpha & -\overline\beta \\ \beta & \overline\alpha 
\end{array} \right),
\qquad  \alpha, \beta \in \mathbf{C}, \quad |\alpha|^2 + |\beta|^2 = 1.
\end{equation}
We show in the appendix A.1 that it coincides with its normalizer and it is not possible to add other connected components.

\item $\mathcal{H}_3^- = SU(1, 1)$,  composed of the matrices of the kind
\begin{equation} 
\left( \begin{array}{cc}
\alpha & \overline\beta \\ \beta & \overline\alpha 
\end{array} \right),
\qquad  \alpha, \beta \in \mathbf{C}, \quad |\alpha|^2 - |\beta|^2 = 1.
\end{equation}
We show in the appendix A.1 that the normalizer is the union of $\mathcal{H}_3^-$ and $i\sigma_2 \mathcal{H}_3^-$. Then we have only to consider the non-connected subgroup
\begin{enumerate}
\item $\mathcal{H}_{3+}^-$, defined by the additional element $i \sigma_2$.
\end{enumerate}

\item $\mathcal{H}_3^0 = E(2)$,  composed of the matrices of the kind
\begin{equation} 
\left( \begin{array}{cc}
\exp(- i\phi) & z \\ 0 & \exp(i\phi)) 
\end{array} \right), 
\qquad  0 \leq \phi < 2 \pi, \quad z \in \mathbf{C}. 
\end{equation}
The normalizer is given by eq.\ (\ref{Triangular}) of the appendix A.3  and the elements of $\mathcal{Q}$ are given by eq.\ (\ref{Q6}). The subgroups obtained by adding other connected components are
\begin{enumerate}
\item $\mathcal{H}_{3,k}^0$ ($k > 1$), defined by the elements (\ref{Q4}).
\end{enumerate}

\item $\mathcal{H}_4^{\infty}$,  composed of the matrices of the kind
\begin{equation} 
\left( \begin{array}{cc}
\exp(t) & is \\ 0 & \exp(- t) 
\end{array} \right),
\qquad  t, s \in \mathbf{R}.  
\end{equation}
Following the procedure indicated in the appendix A.3, we see that the normalizer is composed of the matrices
\begin{equation} 
\left( \begin{array}{cc}
i^{\nu}a & i^{\nu + 1}b \\ 0 & i^{-\nu}a^{-1} 
\end{array} \right),
\qquad \nu = 0, 1, 2, 3, \quad a > 0, \quad b \in \mathbf{R} 
\end{equation}
and the subgroup $\mathcal{Q}$ contains the matrices
\begin{equation}
\left( \begin{array}{cc}
i^{\nu} & 0 \\ 0 & i^{-\nu} 
\end{array} \right),
\qquad \nu = 0, 1, 2, 3.
\end{equation}
The subgroups obtained by adding other connected components are
\begin{enumerate}
\item $\mathcal{H}_{4+}^{\infty}$, defined by the additional element $-e$.
\item $\mathcal{H}_{4++}^{\infty}$, which coincides with the normalizer.
\end{enumerate}

\item $\mathcal{H}_3^{\infty}$, composed of the matrices of the kind
\begin{equation} 
\left( \begin{array}{cc}
\exp(t) & z \\ 0 & \exp(- t) 
\end{array} \right), 
\qquad  t \in \mathbf{R}, \quad z \in \mathbf{C}. 
\end{equation}
The normalizer is given by eq.\ (\ref{Triangular}) of the appendix A.3 and the elements of $\mathcal{Q}$ have the form (\ref{Q8}). The subgroups obtained by adding other connected components are
\begin{enumerate}
\item $\mathcal{H}_{3,n}^{\infty}$ ($n = 2, 3, 4,\ldots$), defined by the elements (\ref{Q2}).
\end{enumerate}
\end{enumerate}

\section{Homogeneous spaces.}

We indicate the homogeneous spaces by the symbol $\Pi$ with the same indices used in the preceding section to label the corresponding stabilizer subgroups. They are defined as quotient spaces, but, in the lower-dimensional cases, it is simpler to consider some of them as orbits in some well known spaces on which $SL(2, \mathbf{C})$ operates linearly.

We consider first the Lie algebra $\mathbf{g} = sl(2, \mathbf{C})$ on which the adjoint representation of $SL(2, \mathbf{C})$  operates. We have seen in section 2 that the orbits composed of semisimple elements have representative elements of the form $\mu M_3 + \nu L_3$. The corresponding stabilizer is defined by the condition
\begin{equation}
h \sigma_3 h^{-1} = \sigma_3,
\end{equation}
namely it is composed of complex diagonal matrices and it is just the subgroup $\mathcal{H}_4$ defined in the preceding section. It follows that all the orbits composed of semisimple elements are isomorphic, as homogeneous spaces, to  $\Pi_4$. 

In a similar way, since the unique orbit composed of nilpotent elements has a representative element of the form $M_1 + L_2$, the corresponding stabilizer is defined by the condition
\begin{equation}
h (\sigma_2 - i \sigma_1) h^{-1} = \sigma_2 - i \sigma_1.
\end{equation}
It follows that
\begin{equation} 
h = \pm \left( \begin{array}{cc}
1 & z \\ 0 & 1 
\end{array} \right),
\qquad  z \in \mathbf{C}, 
\end{equation}
namely the stabilizer is $\mathcal{H}_{4,2}^N$ and the orbit is isomorphic to $\Pi_{4,2}^N$.

The coadjoint representation, which acts on the dual space $\mathbf{g}^*$, is equivalent to the adjoint representation and can be treated in the same way. According to a general theorem \cite{Souriau}, its orbits have an invariant symplectic structure and can play the role of the phase space in some classical Lorentz symmetric Hamiltonian systems.

Then we consider the orbits in the space of four-vectors, which have been classified by Wigner \cite{Wigner}. We use the description of four-vectors by means of Hermitian $2 \times 2$ matrices
\begin{equation}
x = x^k \sigma_k, \qquad \sigma_0 = e.
\end{equation}
The time-like, light-like and space-like orbits have representative elements proportional, respectively, to $e$, $e + \sigma_3$ and $\sigma_3$. The corresponding stabilizers, defined by eq.\ (\ref{Stab}) of the appendix A.1 are,  $\mathcal{H}_3^+$, $\mathcal{H}_3^0$ and $\mathcal{H}_3^-$. It follows that the orbits of time-like, light-like and space-like four-vectors are isomorphic, respectively to $\Pi_3^+$, $\Pi_3^0$ and $\Pi_3^-$. Using a different language, they are, respectively, a sheet of a two-sheet hyperboloid, a light-cone and a one sheet hyperboloid.

Another interesting space is the projective space corresponding the Min\-kowski space, composed of the stright world lines passing through the origin (rays). It can be considered as the velocity space and it contains three orbits. The representative elements of the orbits are the same matrices $x$ listed above, but they are defined up to a constant real factor and the correponding stabilizers are defined by the condition
\begin{equation}
h x h^+ = c x, \qquad h \in \mathcal{H} \qquad c \in \mathbf{R}^*. 
\end{equation}
As we have show in the appendix A.1, if $x = e$ we must have $c = 1$ and $\mathcal{H} = \mathcal{H}_3^+$. If $x = \sigma_3$ we must have $c = \pm 1$ and $\mathcal{H} = \mathcal{H}_{3+}^-$. If $x = e + \sigma_3$ all the positive values of $c$ are allowed and we have $\mathcal{H} = \mathcal{H}_2$. It follows that the orbits in the velocity space are isomorphic, respectively, to $\Pi_3^+$, $\Pi_{3+}^-$ and $\Pi_2$.

From a more geometric point of view, we remark that a time-like ray crosses a sheet of a two-sheet hyperboloid in one point, a light-like ray has a half-line in common with a light cone, and a space-like ray crosses a one-sheet hyperboloid in two points. We see in this way that the orbit of the time-like velocities is isomorphic to a sheet of a two-sheet hyperboloid and the orbit of spacelike-like velocities is isomorphic to a one-sheet hyperboloid with the opposite points identified. The cone formed by the light-like rays intersects a space-like plane on a two-dimensional sphere, which is diffeomorphic to the orbit of the light-like velocities. It is called the {\it celestial sphere} because the position of a star in the sky is characterized by the direction of its light rays detected by an observer. 

We also consider the transitive action of $SL(2, \mathbf{C})$ on the two dimensional complex spinor space. If we choose a representative element, the stabilizer is defined by
\begin{equation} \label{Stabilizer} 
\left( \begin{array}{cc}
a & b \\ c & d
\end{array} \right)
\left( \begin{array}{c}
1 \\ 0 
\end{array} \right) = 
\left( \begin{array}{c}
\alpha \\ 0 
\end{array} \right),
\end{equation}
where $\alpha = 1$. In this way we obtain the subgroup $\mathcal{H}_4^N$ and we see that the homogeneous space  $\mathbf{C}^2$ is isomorphic to $\Pi_4^N$.

$SL(2, \mathbf{C})$ also acts on the projective two dimensional complex space composed of all the complex rays of $\mathbf{C}^2$. The stabilizer is again given by eq.\ (\ref{Stabilizer}), where $\alpha \in \mathbf{C}^*$ and it is $\mathcal{H}_2$. We have obtained another realization of the celestial sphere $\Pi_2$. This is the only two-dimensional homogeneous space and it plays an important role in the construction on the irreducible unitary representations of $SL(2, \mathbf{C})$ \cite{Naimark2,GGV}. 

We can also consider the space of the orbits in the vector space $\mathbf{C}^2$ with respect to the multiplication by arbitrary elements of a given closed subgroup $\mathcal{S}$ of the multiplicative group $\mathcal{C}^*$. $SL(2, \mathbf{C})$ acts on this space and the stabilizer is again given  by eq.\ (\ref{Stabilizer}), where $\alpha \in \mathcal{S}$. It contains the matrices 
\begin{equation} 
\left( \begin{array}{cc}
a & b \\ 0 & a^{-1}
\end{array} \right), \qquad
a \in \mathcal{S}, \quad b \in \mathbf{C}.
\end{equation}

If $\mathcal{S}$ is the group of the complex numbers with modulus one, we obtain again the light-cone $\Pi_3^0$. The relation between the elements of $\mathbf{C}^2$ and the four-vectors of the light-cone is
\begin{equation}
x = 
\left( \begin{array}{c}
z_1 \\ z_2 
\end{array} \right)
\left( \begin{array}{cc}
\overline z_1 & \overline z_2
\end{array} \right).
\end{equation}
Note that the right hand side does not depend on a comon phase factor of $z_1$ and $z_2$.

With other choices of $\mathcal{S}$, we obtain homogeneous spaces isomorphic to $\Pi_3^{\lambda}$, $\Pi_{3,n}^{\lambda}$, $\Pi_3^{\infty}$ and $\Pi_{3,n}^{\infty}$. In all these cases, the orbits of  $\mathbf{C}^2$ cross the three-dimensional sphere
\begin{equation}
|z_1|^2 + |z_2|^2 = 1 
\end{equation}
in one or more points. The homogeneous spaces are compact and diffeomorphic to this sphere, possibly with some points identified. Since the subgroup $SU(2)$ acts transitively on this sphere, it acts transitively also on all the homogeneous spaces listed above.

In conclusion, we have found the following list of homogeneous spaces
\begin{enumerate}

\item $\Pi_6 = SL(2, \mathbf{C})$. The other six-dimensional homogeneous spaces are beyond the scope of the present paper.

\item $\Pi_5^{\lambda}$ ($\lambda \neq 0$). 
\begin{enumerate}
\item $\Pi_{5,n}^{\lambda}$ ($n = 2, 3, 4,\ldots$). 
\item $\Pi_{5,n,\eta}^{\lambda}$ ($n = 2, 4, 6,\cdots, \quad 0 \leq \eta < 1$). 
\end{enumerate}

\item $\Pi_5^0$. 
\begin{enumerate}
\item $\Pi_{5,k}^0$ ($k > 1$).
\item  $\Pi_{5,k,h}^0$ ($k > h \geq 1$).
\item $\Pi_{5,1,h}^0$ ($h > 0$).
\end{enumerate}

\item $\Pi_5^{\infty}$. 
\begin{enumerate}
\item $\Pi_{5,n}^{\infty}$ ($n = 2, 3, 4,\ldots$). 
\item $\Pi_{5,n,\eta}^{\infty}$ ($n = 2, 4, 6,\cdots, \quad 0 \leq \eta < 1$). 
\end{enumerate}

\item $\Pi_5^N$.  
\begin{enumerate} 
\item $\Pi_{5,k,h,\nu}^N$ ($k > 1$, $h \in \mathbf{R}$, $\nu = 0, 1, 2, 3$).
\item $\Pi_{5,k,h,\nu +}^N$ ($k > 1$, $h \in \mathbf{R}$, $\nu = 0, 1$).
\item $\Pi_{5,k,h ++}^N$ ($k > 1$, $h \in \mathbf{R}$).
\item $\Pi_{5,1,h,0}^N$ ($h > 0$).
\item $\Pi_{5,1,h,2}^N$ ($h > 0$).
\item $\Pi_{5,1,h,0+}^N$ ($h \geq 0$).
\item $\Pi_{5,1,h,1+}^N$ ($h > 0$).
\item $\Pi_{5,1,h}^N$ ($h \geq 0$).
\item $\Pi_{5,1,h++}^N$ ($h > 0$).
\end{enumerate}

\item $\Pi_4^N$, the complex two-dimensional spinor space $\mathbf{C}^2$.
\begin{enumerate}
\item $\Pi_{4,n}^N$ ($n = 2, 3, 4,\ldots$). For $n = 2$, we obtain the orbit of the adjoint representation composed of nilpotent elements.  
\item $\Pi_{4,k,n,\eta}^N$ ($k > 1$, $n = 1, 2, 3,\ldots$, $0 \leq \eta < 1$).
\end{enumerate}

\item $\Pi_4$, isomorhic to the orbits of the adjoint representation composed of semisimple elements.
\begin{enumerate}
\item $\Pi_{4,+}$,
\end{enumerate}

\item $\Pi_2$, isomorphic to the celestial sphere.

\item $\Pi_0$, composed of a single point.

\item $\Pi_3^{\lambda}$ ($\lambda \neq 0$), diffeomorphic to the sphere $S_3$.
\begin{enumerate}
\item $\Pi_{3,n}^{\lambda}$ ($n = 2, 3, 4,\ldots$). 
\end{enumerate}

\item $\Pi_3^+$, isomorphic to an orbit of time-like four-vectors namely to a sheet of a two-sheet hyperboloid.

\item $\Pi_3^-$, isomorphic to an orbit of space-like four-vectors, namely to a one-sheet hyperboloid.
\begin{enumerate}
\item $\Pi_{3+}^-$, isomorphic to a one-sheet hyperboloid with the opposite points identified, namely to the orbit of space-like velocities.
\end{enumerate}

\item $\Pi_3^0$, isomorphic to an orbit of light-like four-vectors, namely to a light-cone.
\begin{enumerate}
\item $\Pi_{3,k}^0$ ($k > 1$), obtained from the light-cone by identifying the four-vectors which differ by a factor $k^m$, $m \in \mathbf{Z}$.
\end{enumerate}

\item $\Pi_4^{\infty}$.
\begin{enumerate}
\item $\Pi_{4+}^{\infty}$. 
\item $\Pi_{4++}^{\infty}$.
\end{enumerate}

\item $\Pi_3^{\infty}$, diffeomorphic to the sphere $S_3$.
\begin{enumerate}
\item $\Pi_{3,n}^{\infty}$ ($n = 2, 3, 4,\ldots$).
\end{enumerate}
\end{enumerate}

\section{Covariant mappings.}

Finally we give some remarks on the covariant mappings $\rho: \Pi \to \Pi'$ between two homogeneous spaces. We choose two representative elements $\pi \in \Pi$ and  $\pi' \in \Pi'$ and we indicate by $\mathcal{H}$ and $\mathcal{H}'$ their stabilizers. If we introduce an element $g \in  \mathcal{G}$ with the property $\rho (\pi) =  g \pi'$, we have
\begin{equation} \label{Equi}
g^{-1} \mathcal{H} g \subset \mathcal{H}'.
\end{equation}
Conversely, for every solution $g \in \mathcal{G}$ of this equation, the formula
\begin{equation}
\rho (g' \pi) = g' \rho (\pi) =  g'g \pi', \qquad g' \in \mathcal{G}.
\end{equation}
defines a covariant mapping.

The solutions of eq.\ (\ref{Equi}) can be found by means of elementary calculations on the matrices which represent the group elements. If $g$ is a solution, all the elements of the coset $g \mathcal{H}'$ are solutions and they define the same mapping. In conclusion, a covariant mapping $\rho: \Pi \to \Pi'$ exists if and only if $\mathcal{H}'$ contains a closed subgroup conjugate to $\mathcal{H}$. If we consider the corresponding Lie algebras, a necessary condition is that $\mathbf{h}'$ contains a subalgebra conjugate to $\mathbf{h}$, namely the two subalgebras are connected by an arrow, or by a chain of arrows, in Figure 1. 

If eq.\ (\ref{Equi}) has only one coset of solutions, there is only one covariant mapping. This happens in the following interesting cases.
\renewcommand{\labelenumi}{\alph{enumi})}
\begin{enumerate}
\item If $\mathcal{H} = \mathcal{H}'$, the covariant mapping is an automorphism and the solutions $g$ of eq.\ (\ref{Equi}) belong to the normalizer $\mathcal{N}$. The quotient $\mathcal{N} / \mathcal{H}$ classifies the automorphisms of $\Pi$ and if $\mathcal{N} = \mathcal{H}$ there is only the trivial automorphism.
\item If $\mathcal{H}$ is the connected component of the unit of $\mathcal{H}'$, $\Pi$ is a covering of $\Pi'$ and $\Pi'$ can be obtained from $\Pi$ by identifying some of its points. From eq.\ (\ref{Equi}) it follows that $g^{-1} \mathcal{H} g$ is a connected subgroup of $\mathcal{H}'$, necessarily contained in $\mathcal{H}$. Therefore, $g$ must belong the the normalizer $\mathcal{N}$ of $\mathcal{H}$ and if $\mathcal{N} = \mathcal{H}'$ there is only one covariant mapping of $\mathcal{H}$ into $\mathcal{H}'$, namely the inclusion. If we indicate by $\mathcal{N}'$ the normalizer of $\mathcal{H}'$, one can easily see that $\mathcal{N}' \subset \mathcal{N}= \mathcal{H}'$ and therefore $\mathcal{N}' = \mathcal{H}'$. It follows that there is only one covaraint automorphism of $\mathcal{H}'$.
\item Assume that $\mathcal{H}$ is composed of triangular matrices with the off-diagonal element not depending on the other matrix elements, and that $\mathcal{H}' = \mathcal{H}_2$, the group of all the unimodular triangular complex matrices. As we show in the appendix A.3, a solution  $g$ of eq.\ (\ref{Equi}) must also be a triangular matrix, namely we have $g \in \mathcal{H}'$ and there is only one covariant mapping (the inclusion).
\end{enumerate}

We consider with more detail the physically interesting covariant mappings that connect a three-dimensional homogeneous space (which may describe the energy and the momentum of a particle) and an orbit in the velocity space.  We see from the figure 1 that the only mappings of this kind which are permitted are $\Pi_3^+ \to \Pi_3^+$, $\Pi_3^- \to \Pi_{3+}^-$, $\Pi_{3+}^- \to \Pi_{3+}^-$, while all the other three-dimensional spaces can only be mapped onto $\Pi_2$. As we show in the appendix A.1, in the first case, we are in the situation a) considered just above and in the second and  third case we are in the situation b). In all the other cases we are in the situation c). It follows that that all these mappings are uniquely determined by the Lorentz covariance. 

\section*{Appendices.}
\renewcommand{\thesubsection}{A.\arabic{subsection}}

\subsection{The normalizer of $SU(2)$ and $SU(1, 1)$.}
We consider a subgroup $\mathcal{H} \subset SL(2, \mathbf{C})$ defined by the condition
\begin{equation} \label{Stab}
h x h^+ = x, \qquad h \in \mathcal{H}, \qquad x = x^+ = x^{-1}.
\end{equation}
If $x = e$ (the unit matrix), we obtain the subgroup $SU(2)$ and if $x = \sigma_3$, we obtain the subgroup $SU(1, 1)$. By requiring that $g h g^{-1}$ satisfies the same condition, we obtain after some calculations
\begin{equation}
h (x g^+ x g) = (x g^+ x g) h
\end{equation}
and from the Schur lemma we see that $x g^+ x g$ is a multiple of the unit matrix, namely
\begin{equation}
g x g^+ = c x, \qquad c \in \mathbf{R}^*. 
\end{equation}
If $x = e$, we must have $c = 1$ and we see that $SU(2)$ coincides with its normalizer. If $x = \sigma_3$, we must have $c = \pm 1$ and we see that if  $g$ satisfies this equation with the minus sign, $i\sigma_2 g$ satisfies  the same equation with the plus sign. It follows that the normalizer of  $SU(1, 1)$ is the union of $SU(1, 1)$ and its coset $i\sigma_2 SU(1, 1)$. In this case, $\mathcal{N}/\mathcal{H}$ has only two elements.

\subsection{The normalizer of a subgroup of diagonal matrices.}

We consider a subgroup $\mathcal{H} \subset SL(2, \mathbf{C})$  composed of diagonal matrices and we put
\begin{equation} 
g = \left( \begin{array}{cc}
a & b \\ c & d
\end{array} \right) \in SL(2, \mathbf{C}), \qquad
\left( \begin{array}{cc}
u & 0 \\ 0 & u^{-1}
\end{array} \right) \in \mathcal{H}.
\end{equation}
The matrix
\begin{equation} 
g h g^{-1} =
\left( \begin{array}{cc}
a & b \\ c & d
\end{array} \right)
\left( \begin{array}{cc}
u & 0 \\ 0 & u^{-1}
\end{array} \right)
\left( \begin{array}{cc}
d & -b \\ -c & a
\end{array} \right) 
\end{equation}
must also be diagonal and we see that, if $\mathcal{H}$ contains elements with $u \neq \pm 1$, the normalizer is composed of the matrices
\begin{equation} \label {Normal}
g = \left( \begin{array}{cc}
a & 0 \\ 0 & a^{-1}
\end{array} \right), \qquad
g = \left( \begin{array}{cc}
0 & -a^{-1} \\ a & 0
\end{array} \right), \qquad
a \in \mathbf{C}^*.
\end{equation}

\subsection{The normalizer of a subgroup of triangular matrices.}

We consider a subgroup $\mathcal{H} \subset SL(2, \mathbf{C})$ composed of triangular matrices and we put
\begin{equation} 
g = \left( \begin{array}{cc}
a & b \\ c & d
\end{array} \right) \in SL(2, \mathbf{C}), \qquad
\left( \begin{array}{cc}
u & v \\ 0 & u^{-1}
\end{array} \right) \in \mathcal{H}.
\end{equation}
The matrix
\begin{equation} 
g h g^{-1} =
\left( \begin{array}{cc}
a & b \\ c & d
\end{array} \right)
\left( \begin{array}{cc}
u & v \\ 0 & u^{-1}
\end{array} \right)
\left( \begin{array}{cc}
d & -b \\ -c & a
\end{array} \right) 
\end{equation}
must also be triangular and we obtain the condition
\begin{equation} 
c((u - u^{-1})d - cv) = 0.
\end{equation}
If $v$ can vary independently of $u$, we deduce that $c = 0$, namely the normalizer is composed of triangular matrices, with some other properties which, for various choices of $\mathcal{H}$, can be deduced from the formula
\begin{equation} 
\left( \begin{array}{cc}
a & b \\ 0 & a^{-1}
\end{array} \right)
\left( \begin{array}{cc}
u & v \\ 0 & u^{-1}
\end{array} \right)
\left( \begin{array}{cc}
a^{-1} & -b \\ 0 & a
\end{array} \right) =
\left( \begin{array}{cc}
u & a^2v + ab (u^{-1} -u) \\ 0 & u^{-1}
\end{array} \right).
\end{equation}
When $v$ is an arbitrary complex number, the normalizer is an arbitary triangular unimodular matrix of the kind
\begin{equation} \label{Triangular}
\left( \begin{array}{cc}
a & b \\ 0 & a^{-1} 
\end{array} \right),
\qquad  a \in \mathbf{C}^*,\quad b \in \mathbf{C}.  
\end{equation}

\subsection{Discrete groups of triangular matrices.}

We consider a discrete group $\mathcal{T} \subset SL(2, \mathbf{C})$ composed of triangular matrices and two of its elements
\begin{equation} 
g_1 = \left( \begin{array}{cc}
a_1 & b_1 \\ 0 & a_1^{-1}
\end{array} \right), \qquad
g_2 = \left( \begin{array}{cc}
a_2 & b_2 \\ 0 & a_2^{-1}
\end{array} \right). 
\end{equation}
We have
\begin{equation} 
g(n, m) = g_1^{-n} g_2^m = 
\left( \begin{array}{cc}
A(n, m) & B(n, m) \\ 0 & A(n, m)^{-1}
\end{array} \right), 
\end{equation}
where
\begin{equation} 
A(n, m) = a_1^{-n} a_2^m,
\end{equation}
\begin{displaymath} 
B(n, m) = (a_1^{-n} a_2^m - a_1^{-n} a_2^{-m})(a_2 - a_2^{-1})^{-1} b_2 -
\end{displaymath}
\begin{equation} 
- (a_1^n a_2^{-m} - a_1^{-n} a_2^{-m})(a_1 - a_1^{-1})^{-1} b_1.
\end{equation}

If we assume $a_1, a_2 > 1$, a simple compacity argument shows that it is possible to find two increasing sequences $\{n_i\}$ and $\{m_i\}$ in such a way that
\begin{equation} 
\lim_{i \to \infty} A(n_i, m_i) = C. 
\end{equation}
As a consequence, we also have
\begin{equation} 
\lim_{i \to \infty} B(n_i, m_i) = C (a_2 - a_2^{-1})^{-1} b_2 - C^{-1} (a_1 - a_1^{-1})^{-1} b_1.
\end{equation}
It follows that the limit of the sequence of matrices $\{g(n_i, m_i)\}$ exists and is an element of $SL(2, \mathbf{C})$. Since $\mathcal{T}$ is discrete, this sequence must be stationary for sufficiently large values of $i$ and the sequences $\{A(n_i, m_i)\}$ and $\{B(n_i, m_i)\}$ must have the same property. This is possible only if
\begin{equation} 
(a_2 - a_2^{-1})^{-1} b_2 = (a_1 - a_1^{-1})^{-1} b_1.
\end{equation}

We have shown that if $|a| > 1$ the elements of $\mathcal{T}$ must have the form
\begin{equation} \label{DisTri}
g = \left( \begin{array}{cc}
a & (a - a^{-1})h \\ 0 & a^{-1}
\end{array} \right), 
\end{equation}
where $h$ is a constant which does not depend on $g$. From the group property we see that the elements with  $|a| \leq 1$ have the same form.  In conclusion, the elements of $\mathcal{T}$ are given by eq.\ (\ref{DisTri}), where $h$ is a constant and $a$ belongs to a discrete subgroup of  $\mathbf{C}^*$. The discrete subgroups composed only of elements with $|a|= 1$ have to be treated in a different way.

\section*{Note added.}

After the completion of this work, we have discovered ref.\ \cite{Shaw}, where a classification up to conjugacy of all the subalgebras of the Lorentz Lie algebra is given. Our classification given in section 2 agrees with these preceding results.

\end{document}